# Charge transfer effects, thermo- and photochromism in single crystal CVD synthetic diamond

**R U A Khan**[1], **P M Martineau**[1], **B L Cann**[2], **M E Newton**[2] **and D. J. Twitchen**[3]

[1] Diamond Trading Company, DTC Research Centre, Maidenhead, Berkshire SL6 6JW, UK
[2] Department of Physics, University of Warwick, Coventry CV4 7AL, UK
[3] Element Six Ltd., King's Ride Park, Ascot, Berkshire SL5 8BP, UK

E-mail: riz.khan@dtc.com

**Abstract.** We report on the effects of thermal treatment and ultraviolet irradiation on the point defect concentrations and optical absorption profiles of single crystal CVD synthetic diamond. All thermal treatments were below 850 K, which is lower than the growth temperature and unlikely to result in any structural change. UV-visible absorption spectroscopy measurements showed that upon thermal treatment (823 K), various broad absorption features diminished: an absorption band at 270 nm (used to deduce neutral single substitutional nitrogen ($N_S^0$) concentrations), and also two broad features centred at approximately 360 and 520 nm. Point defect centre concentrations as a function of temperature were also deduced using electron paramagnetic resonance (EPR) spectroscopy. Above ~500 K, we observed a decrease in the concentration of $N_S^0$ centres and a concomitant increase in the negatively charged nitrogen-vacancy-hydrogen complex ($NVH^-$) concentration. Both transitions exhibited an activation energy between 0.6 and 1.2 eV, which is lower than that for the $N_S^0$ donor (~1.7 eV). Finally, it was found that illuminating samples with intense short-wave ultraviolet light recovered the $N_S^0$ concentration and also the 270, 360 and 520 nm absorption features. From these results, we postulate a valence-band mediated charge-transfer process between NVH and single nitrogen centres with an acceptor trap depth for NVH of 0.6-1.2 eV. Because the loss of $N_S^0$ concentration is greater than the increase in $NVH^-$ concentration we also suggest the presence of another unknown acceptor existing at a similar energy as NVH. The extent to which the colour in CVD synthetic diamond is dependent on prior history is discussed.

## 1. Introduction

Chemical vapour deposition (CVD) can be used to produce single crystal diamond of high purity or with controlled doping (Martineau *et al.* 2004, Tallaire *et al.* 2005, Achard *et al.* 2007) and there is interest in using this material for both electronic and optical applications (Isberg *et al.* 2002, Suzuki *et al.* 2004, Koizumi and Suzuki 2006, Trachant *et al.* 2007). The standard production route is by homoepitaxial CVD on {100} substrates processed from synthetic diamond produced using high pressure high temperature (HPHT) synthesis. It is well known that low concentrations of nitrogen in the CVD growth environment can have an important impact on the growth rate, and the nature and concentration of the defects that are incorporated as the material grows (Tallaire *et al.* 2006, Teraji and Ito 2004).

Electron paramagnetic resonance (EPR) studies have helped to elucidate the nature of some of the defects in nitrogen doped CVD diamond and the effect on the optical absorption spectrum is well known in cases such as single substitutional nitrogen ($N_S$) and the nitrogen vacancy (NV) centres. Silicon-containing defects have been well characterized using EPR (e.g. Edmonds *et al.* 2008). Hydrogen-related defects are common in CVD diamond (Fuchs *et al.* 1995), but for defects identified using EPR more recently, such as the negatively charged nitrogen-vacancy-hydrogen and vacancy-hydrogen complexes (Glover *et al.* 2003, 2004) less is known about their influence on optical absorption. The identities of the defects responsible for some of the features in the optical absorption spectrum of nitrogen doped CVD diamond are not yet clear. Simulations have attempted to establish the structure of the NVH centre (Goss *et al.* 2003) and at least in the negatively charged state it appears that the hydrogen is tunnelling between the three equivalent carbon atoms surrounding the vacancy sufficiently quickly that on the timescale of EPR experiments it results in motionally averaged EPR spectra. The absorption spectrum of brown single crystal CVD diamond tends to show a gradual rise in absorption towards shorter wavelengths, often with broad bands at approximately 360 and 520 nm (Martineau *et al.* 2004). It has been suggested that the gradual rise in absorption is caused by vacancy cluster defects (Maki *et al.* 2007) but the defects responsible for the broad absorption bands have not yet been convincingly identified. Similarly, though an absorption feature at 1360 nm (7353 cm$^{-1}$) has been shown (Fuchs *et al.* 1995) to be hydrogen-related using isotopic substitution of deuterium for hydrogen, the nature of the defect that is responsible for this line is not yet clear. However, recently Cruddace *et al.* (2007) correlated the NVH defect with an infrared feature at 3123 cm$^{-1}$.

The ability to monitor a range of different defects in single crystal CVD diamond using EPR has opened up the possibility of using the technique to study thermally activated charge transfer processes. The donor level of $N_S$ in diamond lies approximately 1.7 eV below the conduction band and it was thought that electron donation from $N_s$ would affect the charge states of other defects in a way that could be monitored using EPR. As this line of investigation was developed, it became apparent that the thermally activated charge transfer processes in nitrogen doped single crystal CVD diamond showed a degree of persistence under ambient conditions that suggested the possibility of measuring the parallel changes in the optical absorption spectrum. It also became apparent that the results of studies of such thermochromic effects could be relevant to optical applications of this material. Any temporary change in colour as a result of charge transfer processes induced by heating of CVD synthetic diamond might, for example, potentially affect colour grades given to stones polished for jewellery applications. The existence of thermally activated acceptor states within the bandgap also has relevance to the performance of homoepitaxial CVD diamond as an electronic material.

**2. Experimental**

*2.1 Samples*
The two samples studied in this work were grown on {100} HPHT synthetic diamond substrates in different deposition runs. In both cases, small but significant concentrations of nitrogen were added to the growth environment. Further details on method of deposition are described elsewhere (Martineau *et al.* 2004). After growth the substrates were removed and the samples were polished into free-standing CVD plates on which all subsequent measurements were performed. The dimensions of samples 1 and 2 were 3.4 mm × 3.5 mm × 2.2 mm and 3.7 mm × 3.6 mm × 1.1 mm respectively. Both of these plates possessed a significant degree of colouration: sample 1 was pinkish brown and sample 2 was deep brown.

*2.2 EPR*
Initial concentrations of neutral single substitutional nitrogen ($N_S^0$) and the negatively charged nitrogen vacancy hydrogen complex (NVH$^-$) were determined by electron paramagnetic resonance (EPR) spectroscopy using a Bruker EMX system in combination with an ER041XG-H 90 dB microwave bridge and ER4122 super-high-Q EPR cavity. These measurements were made at room temperature and defect concentrations were determined by comparison to a reference sample (single growth-sector HPHT synthetic diamond doped with 270 ± 5 ppm $N_S^0$) in a standard manner using spectral deconvolution (see for example Tallaire *et al.* 2006).

Isothermal or isochronal annealing was performed *in-situ* with the EPR experiment using a separate system (Bruker EMX spectrometer in combination with an ER041XG 60dB microwave bridge and a $TE_{011}$ cylindrical water cooled high-temperature cavity). In order to perform measurements of defect concentration as a function of temperature, the following procedure was employed: Firstly, the sample was positioned in a quartz dewar (in the high temperature cavity) through which heated nitrogen gas was flowed. The temperature was monitored with a K-type thermocouple, held as close to the sample as possible without perturbing the EPR cavity, the gas temperature was controlled by Oxford Instruments ITC temperature controller whose output was fed to a 1 kW power amplifier to drive the gas heater. Using such an approach, one could heat the sample to 850 K from room temperature in 120 seconds, and cool back down to 550 K in 60 seconds. The sample temperature could be stabilised to within ± 2 K. Prior to heat-treatment, optical illumination was performed on the samples using a high-pressure 200W Hg-Xe arc lamp, exposing them to an unfiltered focussed beam for approximately 5 minutes. The sample was kept in the dark for the rest of the experiment. Following an annealing cycle to a pre-determined temperature for a chosen time, the sample was cooled back to room temperature for the EPR measurements. This procedure was repeated for each temperature up to a maximum of 850 K. The ability to perform annealing and EPR experiments *in-situ* greatly reduced errors resulting from sample positioning within the EPR cavity.

*2.3 Optical absorption*
Optical absorption spectra were measured using a UV-visible-NIR spectrometer (Perkin-Elmer Lambda 9), employing 2 mm apertures and a wavelength range between 200 and 1600 nm. Absorption coefficients in units of $cm^{-1}$ were derived from the measured absorbance and the sample thickness. The samples were then heated to either 798 K or 823 K (525°C or 550°C), employing a ramp rate of 30 K / min, using a hot-stage (Linkam TH1500). The heating conditions were chosen so that the sample would reach the temperature that had been established as being the end-point for the charge transfer process; i.e. when the $N_S^0$ concentrations deduced from the EPR measurements showed no further change. Following heating, the optical absorption characteristics were re-measured using exactly the same conditions as before, ensuring that no light reached the sample prior to measurement.

Next, the samples were illuminated with short wavelength ultraviolet radiation using the xenon arc lamp excitation source of a DiamondView photoluminescence imaging instrument. The duration of illumination was approximately 40 minutes and the sample was rotated every ten minutes to attempt even illumination over all faces. Although use of a 225 nm filter meant that the excitation was predominantly above band gap radiation exciting the near-surface region of the diamond samples, it is known that the filter transmits a significant amount of radiation at wavelengths greater than 225 nm that would penetrate the sample. Finally, the UV-Visible-NIR absorption of the sample was measured, again using exactly the same conditions as those used previously. The following precautions were also employed in order to ensure reproducible optical absorption data: (i) The spectrometer was left turned on for half a day prior to measurement in order to improve its stability. (ii) The samples were carefully aligned in order to ensure maximum signal and therefore good reproducibility between measurements. (iii) All the UV-Visible-NIR absorption measurements were performed on the same day using the same reference spectrum.

Optical micrographs were recorded for both samples following the heat-treatment and ultraviolet illumination steps. A computer-controlled Leica DC300F CCD camera attached to a Wild M420 optical microscope was used to capture the images. One had to ensure that the images were recorded before the illumination light from the microscope obliterated any changes. Therefore the following precautions were taken. The sample changes were performed as rapidly as possible, the period of illumination was reduced to a minimum and all imaging conditions were set initially prior to placing the sample on the stage and these conditions were not changed between the treatment steps.

Fourier transform infrared (FTIR) absorption spectra were measured using a Nicolet Magna IR 750 FTIR spectrometer at a resolution of 0.5 cm$^{-1}$ for both samples before and after thermal treatment and again after UV illumination in order to investigate potential changes in the strength of absorption features.

*2.4 Thermoluminescence*

Thermoluminescence (TL) measurements on the samples were performed as follows. Prior to measurement, the samples were illuminated with ultraviolet light for 30 minutes using the optical excitation lamp and filter of the same DiamondView instrument as was described previously. The TL measurement setup consisted of a photon-counting photomultiplier tube (Hamamatsu R928P) and multi-channel analyzer PCI card (Ortec MCS-PCI) for optical detection, and the same hotstage (Linkam TH1500) as before for sample heating, this time driven using LabView 8.2. The entire setup was light-tight and the measurement was performed under dark ambient conditions in order to minimize background signal. The TL measurement consisted of monitoring the number of counts from the photomultiplier tube over a temperature cycle of between room temperature and 623 K (400°C), employing a ramp rate of 10 K / min. Following measurement a background scan was performed and the background-subtracted signal as a function of temperature was used for the glow curve.

**3. Results**

*3.1 EPR measurements*

EPR spectroscopy, being a quantitative technique, was used to investigate the concentration of the $N_S^0$ and $NVH^-$ centres as a function of annealing temperature, those being the EPR active defects with highest concentrations. Other characteristic CVD diamond defects were observed in much lower concentrations such as not to greatly affect the overall charge balance, and hence these are not discussed further. The variation in concentration of these point defects as a function of temperature is shown in figure 1(a) for sample 1 and (b) for sample 2.

We first observe that sample 1 displays a reduction in $N_S^0$ and a corresponding rise in $NVH^-$ above a transition temperature of 600 K. Overall the $N_S^0$ concentration decreased by 180 ppb, whilst the $NVH^-$ concentration increased by 70 ppb. For sample 2, we also observe a significant decrease in the $N_S^0$ concentration and a concomitant rise in the $NVH^-$ concentration at a temperature of 500 K. Overall the $N_S^0$ concentration decreased by 1570 ppb, whilst the $NVH^-$ concentration increased by 610 ppb. The solid curves in figure 1 show best-fits to the experimental isochronal annealing data using second order kinetics, but for these data sets simulations with first order kinetics produce equally good fits. For sample 1, we extracted an activation energy for the loss in $N_S^0$ and gain in $NVH^-$ of 1.1(2) eV and for sample 2 the best fit was obtained with an activation energy of 0.6(2) eV. Studies of six other samples (not discussed) produced activation energies in this range. The effect of thermal treatment on the EPR spectrum was found to be reversed by UV illumination.

One HPHT synthetic nitrogen doped diamond was measured in a similar manner. The difference in measured $N_S^0$ concentration over this thermal treatment range was less than 2%, and therefore we conclude that treatment up to 850K does not lead to any of the changes that were observed in the CVD synthetic samples.

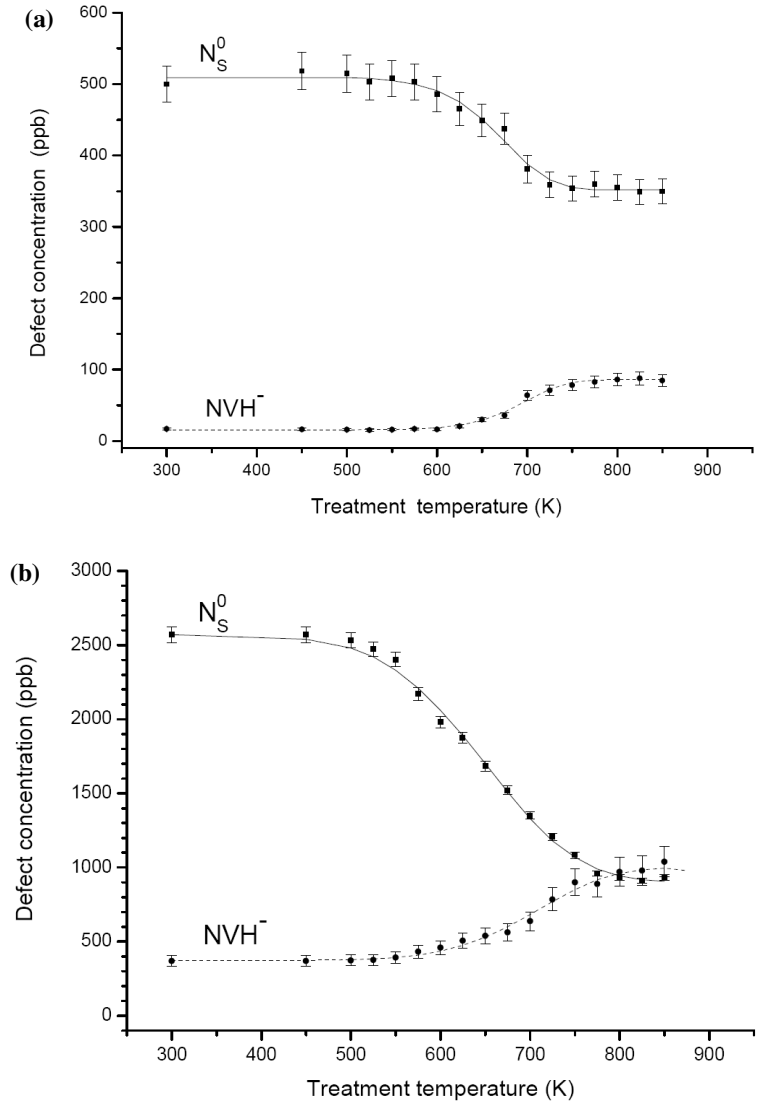

**Figure 1.** The concentrations of $N_S^0$ and $NVH^-$ defect centres in (a) sample 1 and (b) sample 2, measured at room temperature by EPR as a function of isochronal (8 minutes) annealing temperature. The sample was kept in the dark during the annealing and measurement cycles. Prior to the first anneal the sample was illuminated with unfiltered light from a 200 W HgXe arc lamp for 5 minutes.

*3.2 UV-visible spectroscopy measurements*
Figure 2 shows room-temperature UV-visible optical absorption spectra for sample 1 (A) in its initial as-grown state, (B) following heating to 798 K, and (C) after ultraviolet illumination. In (A) and (C), strong absorption features at approximately 270, 360 and 520 nm are observed. The feature at approximately 270 nm corresponds to $N_S^0$ centres (Chrenko *et al.* 1971) and the broad absorption features centred at approximately 360 and 520 nm are commonly observed in nitrogen-doped CVD synthetic diamond but their origin is still a matter of debate (Martineau *et al.* 2004). The optical absorption profile shows a general increase in magnitude for progressively shorter wavelengths. We term this "ramp" absorption and the strengths of this and the 520 nm absorption feature strongly influence the perceived colour of these brown CVD diamond samples. The strength of the 360 nm band may also affect the colour because it is broad enough to extend into the visible region of the spectrum.

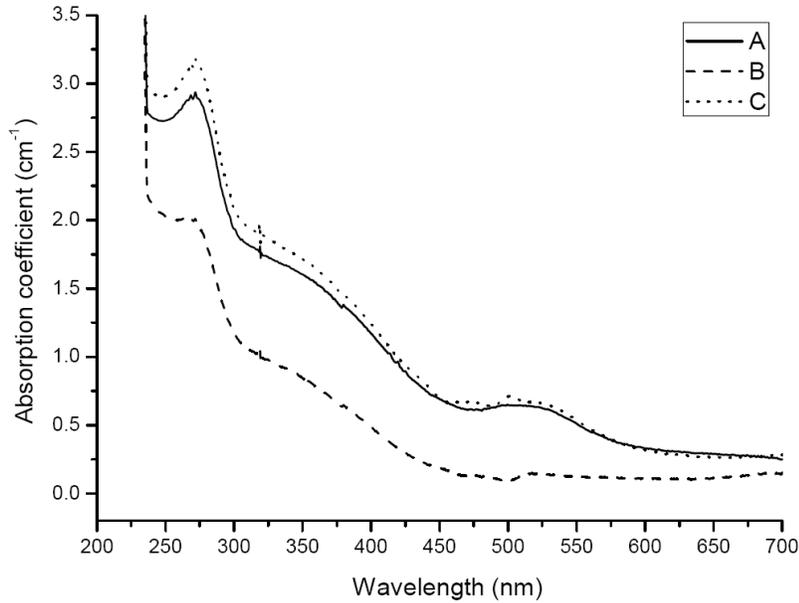

**Figure 2.** Room-temperature UV-visible absorption data between 200 and 700 nm, showing absorption coefficients as a function of wavelength for sample 1 in its initial as-deposited state (A, solid line), following heating up to 798 K (ramp rate 30 K / min) (B, dashed line) and following exposure to ultraviolet irradiation for a duration of 40 minutes (C, dotted line). Absorption coefficients were deduced by normalising with the sample thickness.

Comparing the plots, we observe that spectrum (A) and (C) are qualitatively similar but in spectrum (B) the 270 nm, 360 nm and 520 nm absorption bands are much weaker. Comparing (C) with (A), we can observe that the 270 and 360 nm absorption features appear to be slightly stronger in (C). Overall, we can conclude that the heating procedure has weakened the three main absorption features and that the illumination step has returned the absorption profile to a similar state to what it was originally.

Qualitatively similar results were obtained for sample 2, though the effects are greater. Figure 3 shows the optical absorption spectra for this sample in its initial state, after heat treatment and finally after subsequent UV illumination. We note that the heating stage has almost completely removed the 270 nm feature, and that it has strengthened markedly upon ultraviolet irradiation. Fitting to a spectrum of type Ib synthetic diamond of known nitrogen content (De Weerdt and Collins 2008), we deduce a change in the $N_S^0$ concentration from 0.5 to 3.5 ppm. Also, as previously we notice that the 360 and 520 nm features have weakened markedly upon heating and strengthened upon UV illumination. We also observe the existence of another very broad absorption feature centred around 900 nm. This band shows a behaviour which is the reverse of the other features, being stronger after thermal treatment and weaker following ultraviolet illumination. Finally we observe a sharp, narrow absorption feature at approximately 1360 nm. It is known that a hydrogen-related defect is responsible for this line. Like the broad feature at 900 nm, the 1360 nm line has increased in strength after thermal treatment and weakened after UV illumination.

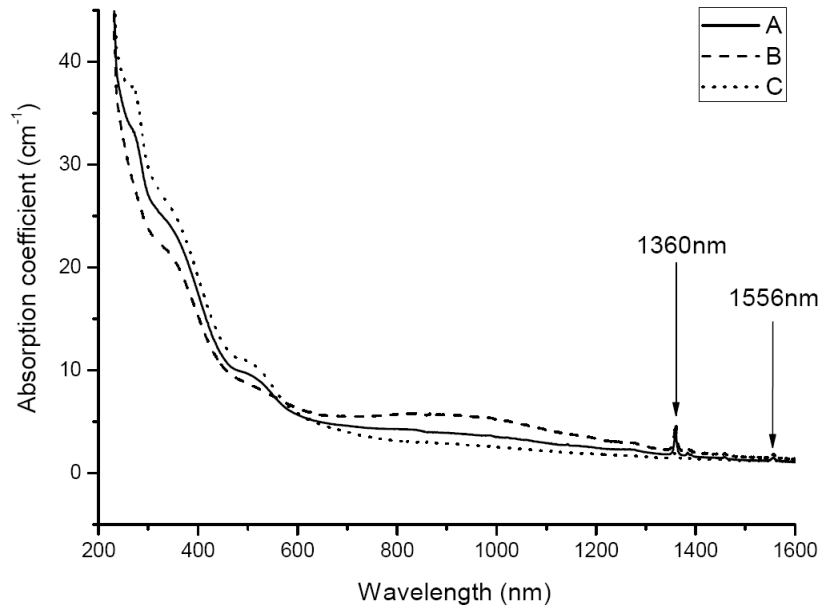

**Figure 3.** Room-temperature UV-visible-near infrared absorption data between 200 and 1600 nm, showing absorption coefficients as a function of wavelength for sample 2 in its initial as-deposited state (A, solid line), following heating up to 823 K (ramp rate 30 K / min) (B, dashed line) and following exposure to ultraviolet irradiation for a duration of 40 minutes (C, dotted line). Absorption coefficients were deduced by normalising with the sample thickness.

In order to highlight changes that are not clearly apparent in original data plots we show UV-visible difference spectra in figure 4. The results are plotted so that increasing positive y-values correspond to increased absorption after the UV illumination stage, and decreasing negative y-values correspond to increased absorption after the thermal treatment. As commented upon previously we observe in both samples that the strength of the 270 nm absorption feature has increased markedly upon UV illumination, as has the strength of both the 360 and 520 nm absorption features. Indeed these absorption features are now clearly visible as distinct features in the difference spectra for both samples. There also appears to have been an increase in 'ramp' absorption, which is the overall absorption background that gradually increases with decreasing wavelength. One may also observe that the changes in absorption coefficient are much greater in sample 2 than in sample 1. In sample 2, we can more clearly observe the increase in the very broad absorption feature at approximately 900 nm. This feature is also present in sample 1 but it is much weaker, and therefore the spectrum in this region for this sample possesses some spectrometer artefacts. Finally, in both samples we can observe that after the heat treatment, the strength of various sharp absorption features at 1360, 1458 and 1550 nm has increased.

Charge transfer effects, thermo- and photochromism in single crystal CVD synthetic diamond

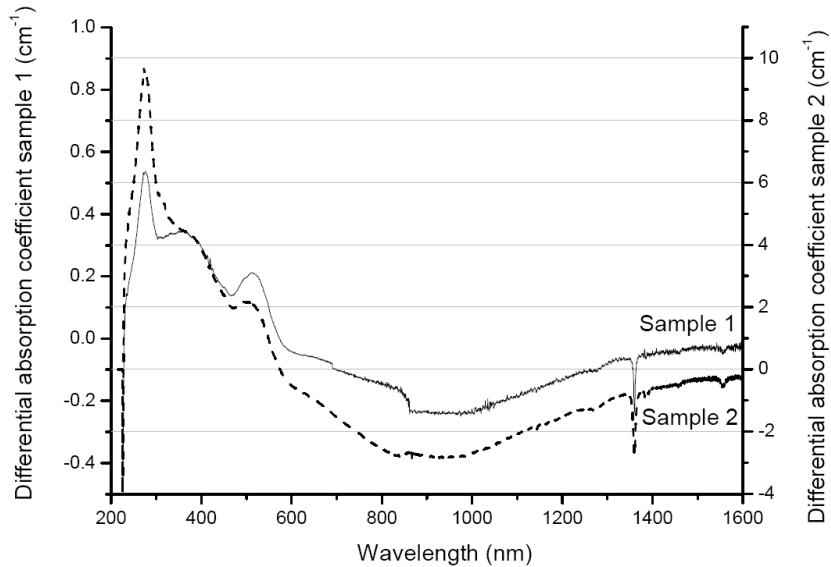

**Figure 4.** UV-visible-NIR absorption difference spectra for sample 1 (solid line, left scale) and sample 2 (dashed line, right scale), deduced by subtracting the optical absorption coefficient values as a function of wavelength measured after ultraviolet illumination (40 minutes) with those values measured after heating (798 or 823 K). More positive values indicate that a feature has increased in strength after illumination, and more negative values indicate that it has increased after heating.

*3.3 Optical micrographs*
The magnitude of these changes recorded is easily sufficient to observe with the naked eye. Evidence for this is provided in figure 5 which shows optical micrographs of both sample 1 (Fig. 5(a) and (b)) and sample 2 (Fig. 5(c) and (d)). (a) and (c) correspond to images of the samples after the heating step and (b) and (d) correspond to images of the samples after the UV illumination step. We can see that sample 1 is fairly colourless after the heat treatment step and is pink following UV illumination. In sample 2 we note that the sample has become redder and the colour is more saturated following UV illumination. CIE colour coordinates for standard lighting conditions may be derived from absorbance spectra (for an outline of the method, see Twitchen *et al.* 2004) and, when this procedure is applied to the spectra measured for these samples, the colour changes following UV illumination correspond to a large increase in *a\** (red) and a smaller increase in *b\** (yellow) following UV illumination.

*3.4 FTIR absorption spectroscopy*
We noted that for both samples thermal treatment to 800 K resulted in reduction in the strength of the line at 1344 $cm^{-1}$ that is attributed to $N_S^0$ and an increase in that of the line at 1332 $cm^{-1}$, to which $N^+$ is known to be a contributor (Lawson *et al.* 1998). A line at 3123 $cm^{-1}$, observed in the spectrum of both samples in their as-grown state, was not observed after thermal treatment to 850 K. After UV illumination the 3123 $cm^{-1}$ line was restored.

*3.4 Thermoluminescence measurements*
Thermoluminescence measurements were recorded between room temperature and 673 K for both samples using the method described earlier. Figure 6 shows TL the glow curves for the two samples. We can observe in both cases a strong and fairly sharp TL peak at around 630-650 K. This feature appeared to be stronger in sample 1 but this might also be accounted for by differences in optical coupling. For both samples an activation energy of 1.2 ± 0.2 eV was deduced for this TL peak. In Sample 1 there was also a weak feature at 330-360 K. We speculate that this low temperature feature could be due to donor-acceptor luminescence because of very small concentrations of boron impurities (the B acceptor level is 0.37 eV (Freitas *et al.* 1994) above the valence band edge and electrons can therefore thermally excite from the valence band to B acceptors at these temperatures).

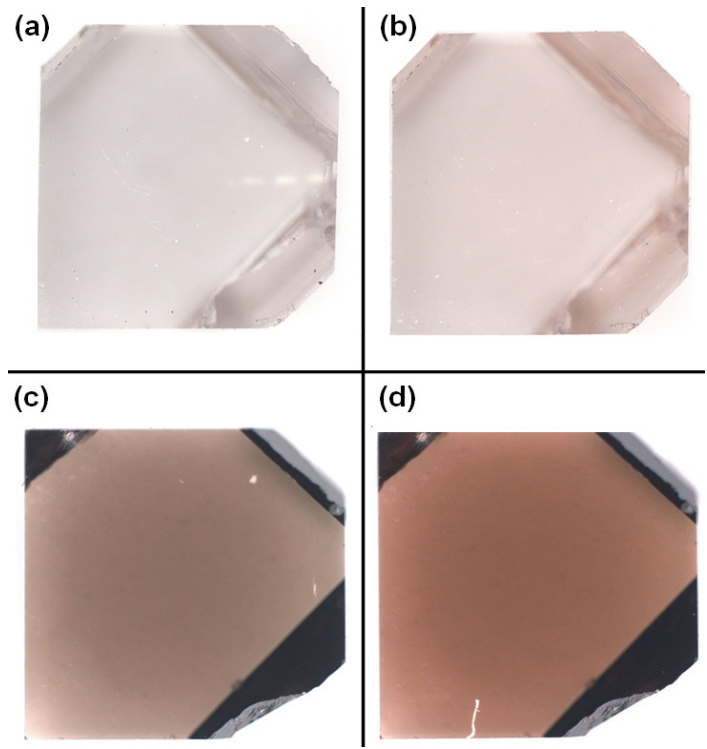

**Figure 5.** Optical micrographs of both samples. The two images at the top correspond to sample 1 following heating to 798 K (a) and following 40 minutes' ultraviolet irradiation (b). The images at the bottom correspond to sample 2 following heating to 823 K (c) and following 40 minutes' ultraviolet irradiation (d). All micrographs were taken at room temperature using a transmission microscope.

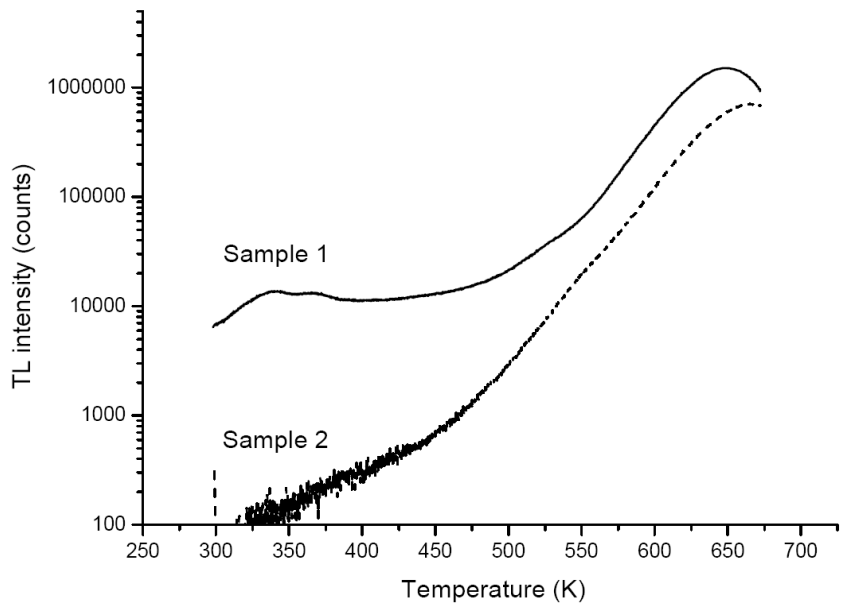

**Figure 6.** Thermoluminescence glow-curves of both samples, taken between room temperature and 673 K. Prior to measurement both samples were illuminated with ultraviolet light for 30 minutes. The samples were kept in the dark for the duration of the heating cycle.

## 4. Analysis of UV/visible absorption spectra

In order to deduce the relative variations in strength of the various absorption features caused by thermal treatment and UV illumination, we synthesized the absorption spectra of sample 1 (shown in figure 2) by summing the various components to create a fit to the measured spectra. As discussed previously we fit the 270 nm absorption feature in order to ascertain the concentration of $N_S^0$. Likewise, absorption fitting was performed on the 360 and 520 nm features in order to deduce their relative strength. The resulting data is shown in table 1. We see that the strengths of the 270, 360 and 520 nm absorption features have diminished following the heating step. The reduction in strength of the 270 nm feature is attributable to a decrease in the concentration of $N_S^0$ centres from approximately 0.5 to 0.3 ppm in agreement with EPR concentration measurements (we estimate errors of ±0.1 ppm in these values).

We also attempted to fit the difference spectra for both samples (shown in figure 4) to type I$b$ + 360 nm + 520 nm + ramp absorption spectra, over the 230 - 800 nm region of the spectrum, in order to deduce whether only these absorption features were responsible for the variations. The ramp constant, $R$, is defined by the function $\alpha_{ramp} = R / \lambda^3$ where $\alpha_{ramp}$ is the ramp contribution to the overall absorption coefficient and $\lambda$ is the wavelength in nm. For both samples moderately good fits were achieved without the need for additional features or changes in the width of the observed absorption features. The results are shown in table 2. For sample 1, all of the absorption changes could be accounted for without any change in ramp absorption. For sample 2, the best fit suggests that the ramp absorption has changed significantly.

**Table 1.** Fit parameters for the optical absorption spectra of sample 1 under the three cases (A) before treatment, (B) following heating and (C) following ultraviolet irradiation, showing the concentration of $N_S^0$ centres derived using the 270 nm absorption peak and the relative intensities of the 360 and 520 nm absorption bands.

|  | A | B | C |
|---|---|---|---|
| $N_S^0$ [270 nm band] (ppm) | 0.50 | 0.30 | 0.65 |
| 360 nm band (cm$^{-1}$) | 0.4 | 0.2 | 0.5 |
| 520 nm band (cm$^{-1}$) | 0.3 | 0.1 | 0.3 |

**Table 2.** Fitting of the UV-visible-NIR absorption difference spectra shown in figure 4 to (i) a ramp that increases with decreasing wavelength, (ii) 270 nm absorption ($N_S^0$) and absorption bands centered at (iii) 360 and (iv) 520 nm. Note that (i) is in arbitrary units. Please see text for an explanation of the ramp constant, $R$.

|  | Sample 1 | Sample 2 |
|---|---|---|
| Ramp constant, $R$ | 0 | 4 x 10$^7$ |
| $\Delta N_S^0$ [270 nm band] (ppm) | 0.15 | 2.3 |
| 360 nm band (cm$^{-1}$) | 0.20 | 1.5 |
| 520 nm band (cm$^{-1}$) | 0.18 | 1.5 |

## 5. Discussion

$N_S^0$ is an important electron donor in these samples. Standard mechanisms for charge transfer from $N_S^0$ to another defect involve thermal or optical excitation of an electron from $N_S^0$ into the conduction band and subsequent capture of an electron at the second defect. It is clear, however, that the activation energies deduced from the EPR and TL data are significantly less than the 1.7 eV required to excite an electron from $N_S^0$ to the conduction band. An alternative mechanism involves excitation of a valence band electron to an acceptor defect, with the resulting hole in the valence band being filled as a result of loss of an electron from $N_S^0$.

The concentration of $NVH^-$ is seen to increase as the concentration of $N_S^0$ decreases, strongly suggesting that at least some of the charge from $N_S^0$ centres is transferred to NVH defects with an acceptor level approximately lying between 0.6 and 1.2 eV above the valence band. EPR data indicates, however, that for both samples the decrease in the concentration of $N_S^0$ is approximately 2.5 times the increase in that of $NVH^-$. This requires the existence of at least one other kind of defect that captures electrons while not being observed in the EPR spectrum in either their initial or final charge states. We give such defects the label "X".

Our conclusions from the EPR results can be summarized thus: (i) an NVH acceptor level exists between 0.6 and 1.2 eV above the valence band edge and (ii) at least one other defect "X" has an acceptor level at a similar energy to NVH. Figure 7 shows a band diagram illustrating the defect levels involved. The two processes that explain the observations are given by equations 1 and 2, with thermal treatment and UV illumination driving the processes to the right and left sides of the equations, respectively.

$$N_S^0 + NVH^0 \leftrightarrow N_S^+ + NVH^- \qquad (1)$$

$$N_S^0 + X \leftrightarrow N_S^+ + X^- \qquad (2)$$

The FTIR results are consistent with this interpretation. Thermal treatment resulted in reduction in the strength of the 1344 cm$^{-1}$ line attributed to $N_S^0$ and an increase in that of the 1332 cm$^{-1}$ line, to which $N_S^+$ is known to be a contributor (Lawson *et al.* 1998). For measurements on a large number of N-doped CVD diamond samples, Cruddace *et al.* (2007) also noted a reasonable correlation between the strength of a line at 3123 cm$^{-1}$ and the concentration of $NVH^-$ measured using EPR, and therefore proposed that the line corresponded to the $NVH^-$ centre. However in the present study we observed that after thermal excitation which causes $NVH^0$ to be converted to $NVH^-$, the 3123 cm$^{-1}$ line disappeared. We therefore propose that this line is in fact a vibrational mode of $NVH^0$ rather than $NVH^-$, and that equivalent mode of $NVH^-$ is significantly shifted from this region of the spectrum and has yet to be identified.

We now move our attention to the UV/visible/NIR absorption data for the two samples. The most obvious link with the EPR data is the observation that the 270 nm $N_S^0$-related feature was reduced by the thermal treatment in the same way observed for the EPR lines attributable to $N_S^0$. Although the 270 nm feature is in the ultraviolet region of the spectrum, $N_S^0$ also causes absorption as far into the visible region of the spectrum as 500 nm, and therefore, even by itself, the change in $N_S^0$ induced by the thermal treatment would influence the perceived colour of the sample. It clear, however, that the defects responsible for the ramp feature and the two broad bands centred at approximately 360 nm and 520 nm are also involved in thermally activated charge transfer processes that have a significant influence on the colour of the samples. The strengths of the 360 and 520 nm bands are reduced by the same thermal treatment that reduces the $N_S^0$ concentration and increased by UV illumination that increases the $N_S^0$ concentration and, as indicated by equations 1 and 2, this suggests that they are attributable to defects such as X and $NVH^0$ that accept electrons during thermal treatment and are returned to their original charge state by UV illumination. The ramp in the absorption spectrum shows similar behaviour to the 360 and 520 nm absorption bands for sample 2 but for sample 1 no significant change in the ramp absorption was observed.

The strengths of the absorption features at 1360 nm and approximately 900 nm are increased by the thermal treatment but reduced by UV illumination, suggesting that the defects responsible are produced by capture of electrons during thermal treatment and change charge state by loss of electrons during UV illumination. The changes in the 1360 cm$^{-1}$ line intensity were reversible upon UV illumination and thermal treatment which indicates that no structural changes in the defect responsible for this feature were occurring. The response of these absorption features to thermal treatment and UV illumination is the reverse of what we have observed for NVH$^0$. Such behaviour is what would be expected for defects such as NVH$^-$ but, for the 900 nm band in any case, more work will be required to make a definitive assignment. The 1360 nm line is unlikely to correspond to the NVH$^-$ defect for the following reason: Although the 1360 nm and NVH$^-$ defects anneal at approximately the same temperature, in samples grown on (100) oriented substrates when the absorption spectra are recorded with light linearly polarized either parallel or perpendicular to the growth direction, the former can exhibit marked preferential orientation with respect to the growth direction (to be published). In the samples showing this preferential orientation for the optical system the NVH$^-$ defect (with motionally averaged trigonal symmetry) shows no preferential orientation with respect to the growth direction.

We have also considered the possibility that NVH$^0$ defects are responsible for the 520 nm band. Studies of the effect of higher temperature annealing suggest that NVH$^-$ (detected using EPR), the 3123 cm$^{-1}$ absorption line (which we have tentatively attributed to NVH$^0$) and the 520 nm absorption band all anneal out in the same temperature range (approximately 2000-2200 K). As discussed above, however, the results of this study suggest that the acceptor level of NVH is only approximately 0.6 - 1.2 eV above the valence band edge and it is therefore not obvious to us why an optical absorption feature at 520 nm (~2 eV) should be observed.

In other investigations of brown single crystal CVD diamond, we noted a general correlation between the strength of the band at approximately 360 nm and the ramp absorption feature. It is not known what defect or defects are responsible for these two features but it has been suggested that the ramp in the absorption spectrum is caused by vacancy cluster defects (Jones 2008) that have internal surfaces with bonding that shows sp$^2$-like characteristics.

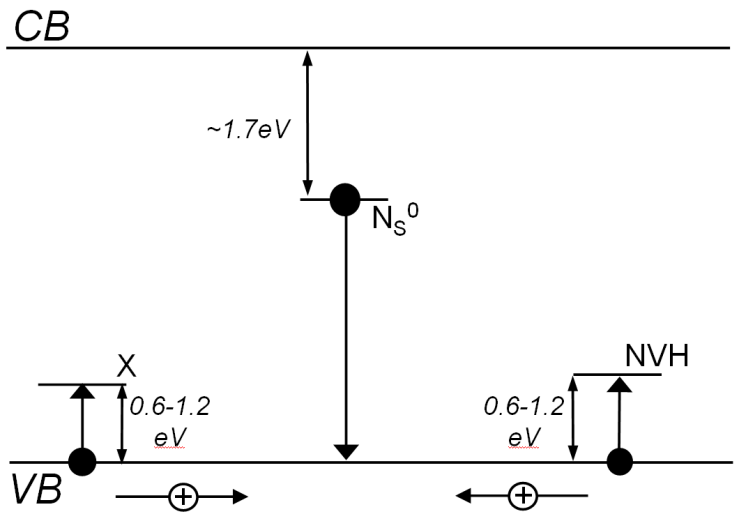

**Figure 7.** Proposed band diagram, showing the thermal release of electrons within the valence band to acceptor levels of known (NVH) and unknown (X) origin that exist between 0.6 and 1.2 eV from the band edge, and recombination between neutral single substitutional nitrogen and the holes that are created in the initial step.

The observation of a strong thermoluminescence feature at ~620-670 K indicates that carriers are activated from trap states at the same temperatures at which EPR and optical absorption spectroscopy show that charge transfer processes occur. It is therefore very likely that this strong TL feature is generated by thermal excitation of electrons from the valence band to either NVH or X acceptor levels. Both lead to the formation of holes in the valence band. This suggests that the observed TL feature is a result of recombination of electrons from $N_S^0$ centres and the thermally-created holes in the valence band. Further measurements (e.g. TL spectroscopy) will be performed in order to elucidate the process and to show whether the transition is direct or involves donor-acceptor recombination.

There may be significant heterogeneity in defect distributions within one CVD diamond sample. Luminescence images acquired using DiamondView $^{TM}$ have shown striations in nitrogen-vacancy luminescence which is attributed to the differential uptake of impurities on growth terraces and risers (Martineau *et al.* 2004). Also small variations in growth conditions, for example temperature, may lead to changes in the impurity concentrations. The $N_s^0$ and $NVH^-$ values deduced through these EPR measurements are bulk values and take no account of variations within a sample.

The colour of some rare natural diamonds is also affected by thermal treatment and UV illumination. The colour of very rare 'chameleon' stones (Fritsch *et al.* 2007) generally changes from a stable 'olive' green colour to an unstable yellow due to alterations in the strength of two broad absorption features: one at approximately 480 nm and the other centred at around 800 nm. These stones tend to show trace concentrations of $N_S^0$ centres (de Weerdt and van Royen 2001) and they often exhibit phosphorescence, which is an indicator of charge transfer. Pink type IIa natural diamonds also form another important class of diamonds that exhibit colour changes. These have been shown to change colour from pink to brown upon UV illumination (de Weerdt and van Royen 2001). The pink colour is a result of a broad optical absorption feature at 550 nm that is bleached by ultraviolet radiation and also diminishes on cooling. It therefore appears that, although the mechanisms have not yet been fully resolved, charge transfer effects in natural diamonds can lead to colour changes. Most nitrogen in natural diamond is aggregated but, at least in some cases, it is possible that low concentrations of single substitutional nitrogen play a role in these charge transfer processes.

**6. Conclusion**

These results have shown that some single crystal CVD synthetic diamond may exhibit optical absorption features that are not stable. Either annealing at moderate temperatures (~800 K) or bright illumination may modify the strength of these features. These effects are entirely reversible, i.e. UV illumination will return a feature diminished by thermal treatment, and vice versa. We have discussed our results within the context of a charge transfer model between acceptor defects and single substitutional nitrogen and shown how thermoluminescence data can be accommodated by such a model. We also note that these processes are temporally stable as long as the charge populations are not disturbed by incident light (or heating). As a result of these effects, not only broad optical absorption features that contribute to colour, but also narrow lines both in the UV-visible and IR spectrum (corresponding to point defect centres) are influenced. The analysis of point defect concentrations should therefore be approached with caution if the prior history of the sample is not known. The specific details of the effects reported here depend on the presence mixtures of defects (such as NVH) which have not been observed in other kinds of diamond. Single substitutional nitrogen, which is rarely found in significant concentrations in natural diamond, acts as a source of electrons and therefore drives the charge transfer process.

This work has implications both for fancy-coloured and near-colourless CVD synthetic diamond material. The colour changes are significant and may correspond to as much as several GIA colour grades (for an explanation of GIA's grading system, see e.g. King *et al.* 1994). Although these experiments have employed conditions that would be unusual to encounter in a standard diamond grading environment, some colour changes are likely under normal grading conditions. Standard tests applied to diamonds (such as examination in DiamondView$^{TM}$) involve illumination with short wavelength UV radiation which we have shown may affect the perceived colour. It is also true that the temperature of stones may on occasion, for example during mending of jewellery, be raised high enough to cause a temporary colour change. Further work will be directed towards the study of the effect across a greater range of samples, developing a better understanding of the thermoluminescence process, and also to investigate the electronic processes that occur in the illumination step, for example by study of the dependence on excitation energy.

**Acknowledgements**
The authors would like to acknowledge A. Bennett, H. Dhillon and S. Woollard (E6) for growing some of these samples, and A. Taylor, S. Sibley, C. Kelly and P. Leno (DTC) for assistance with sample characterisation.